\documentclass[11pt]{article}
\usepackage[T1]{fontenc}
\usepackage{lmodern}
\usepackage[margin=1in]{geometry}
\usepackage{microtype,amsmath,amssymb,graphicx,booktabs,array,placeins,float}
\usepackage{xcolor}
\usepackage[colorlinks=true,linkcolor=blue!50!black,citecolor=blue!50!black,urlcolor=blue!50!black]{hyperref}
\hypersetup{pdftitle={When AI Advisors Agree: Distinguishing Forecast Risk from Forecast Reliability},pdfauthor={Russlan Ramdowar}}
\title{When AI Advisors Agree:\\Distinguishing Forecast Risk from Forecast Reliability}
\author{Russlan Ramdowar\\\small Future Edge Group FZE, United Arab Emirates\\\small iPulse AI\\\small\texttt{russlan@ipulseai.com}\\\small ORCID: \href{https://orcid.org/0009-0004-9311-6957}{0009-0004-9311-6957}}
\date{14 October 2026 (version 2)}
\begin{document}
\maketitle
\begin{abstract}
If seven AI advisors agree, what have we learned: that their forecast is reliable, or that the asset is easier to forecast? We examine 928 completed first-quarter outcomes for 232 assets across four historical iPulse AI batches. Ten Gemini configurations form seven equally weighted investment frameworks; no forecasts are regenerated. The council has mean absolute return error of 15.15 percentage points (pp), versus 14.99 for no change, with an unresolved paired difference of +0.16 pp. Error rises from 10.31 to 21.05 pp across disagreement quartiles, but the no-change baseline rises almost identically, from 9.89 to 21.15. In the pooled archive, selecting the lowest-volatility quarter produces lower error than selecting the least-disagreeing quarter: 7.97 versus 10.31 pp. Directional agreement introduces a second trap: its apparent value depends on the aggregation rule. When at least six of seven frameworks share a direction, the sign of their mean return is correct in 53.21\% of cases, versus 44.11\% with lower agreement. Scoring the modal direction instead changes the lower-agreement accuracy to 54.27\%, removing that positive contrast. Recomputed-rank bootstraps, scale-aware losses, and period sensitivities qualify these descriptive results; four overlapping periods cannot establish future-regime performance. Agreement, visible asset risk, and aggregation-specific skill should be evaluated separately before consensus is presented as confidence.
\end{abstract}

\section{Introduction}
An AI council can produce a persuasive form of evidence: several advisors, different reasoning frameworks, and a common conclusion. Agreement is visible immediately; correctness arrives later. This gap makes consensus attractive as a confidence signal, especially in forecasting tasks where the answer does not yet exist. It also makes agreement easy to overinterpret.

Consider two assets. One has a stable price history, and seven advisors predict modest returns close to one another. The other is volatile, and their predictions differ widely. If the first council later has smaller errors, agreement appears to predict reliability. But a forecast that ignores the council and simply predicts no price change might have smaller errors on the first asset too. The council could be identifying \emph{forecast difficulty}, rather than identifying when its own judgment adds value.

This paper tests that distinction in the ordinary production archive of iPulse AI, an Open Agentic Investment Research Platform. Its research pipeline generates reports through configured investment frameworks and modes, followed by a synthesis layer. The present study isolates a transparent, equally weighted council from historical individual forecasts. It does not evaluate the evidence-weighted synthesizer, rank current provider models, or measure the profitability of an investment strategy. A separate study of this platform examines how model disclosure affects synthesis weights \cite{ramdowar2026}; here the question concerns subsequent realized outcomes.

We analyze four batches issued between April and July 2026. The central comparison asks whether forecast dispersion predicts error \emph{beyond} what is already visible in an asset's price variability and in a no-change baseline. We then distinguish agreement about numerical returns from agreement about up, flat, or down. These are related signals, but they answer different questions.

We use \emph{reliability} as a question to be tested, not a single measured property. Raw return error measures numerical precision; error relative to a baseline measures incremental skill; directional accuracy measures up/flat/down classification. None establishes probabilistic calibration. The contribution is a historical outcome study with three connected results: the disagreement--error gradient is largely mirrored by a no-change baseline; historical volatility selects lower-error forecasts than raw disagreement; and the apparent directional benefit of agreement depends on how framework forecasts are aggregated. Public forecasts and aggregate diagnostics support inspection of these claims, while full outcome replication requires compatible authorized market inputs.

\section{Related work}
The distinction between disagreement and uncertainty predates language models. Lahiri and Sheng \cite{lahiri2010} show, under their forecasting model, that dispersion among forecasters does not capture all uncertainty associated with future common shocks. Glas \cite{glas2020} documents that the disagreement--uncertainty relationship depends on the forecast setting and how the quantities are measured. Our study brings this concern into an operational AI council: a forecast-spread signal should be tested against ordinary, observable task risk before it is interpreted as model-specific confidence.

In machine learning, deep ensembles offer a practical route to uncertainty estimation \cite{lakshminarayanan2017}. Those results do not establish that prompt frameworks sharing a model backbone behave like independently trained ensemble members. For language models, Xiao et al.\ \cite{xiao2025} formalize and test relationships between generation consistency and correctness across several tasks. Their work motivates examining consistency empirically rather than assuming that it is a universal confidence measure. Our task differs in an important respect: answers are later market returns, rather than labels already available when generations are evaluated.

ForecastBench \cite{karger2024} emphasizes forecasts about genuinely future events. We likewise preserve the distinction between the generation date and the outcome date, but use archived production returns rather than a purpose-built question benchmark. This observational setting provides breadth at the cost of weaker control over the information supplied to each historical run.

Selective prediction studies the tradeoff between retaining more predictions and accepting more error \cite{elyaniv2010}. We use retention curves descriptively: how much error remains when a rule keeps only a specified fraction of forecasts? We do not transfer selective-classification guarantees to market returns. Finally, point-forecast evaluation requires attention to the forecast functional and scoring rule \cite{gneiting2011}. Our arithmetic council mean is evaluated under absolute error as a practical decision loss; it is not asserted to be an optimal conditional median. A median-aggregation sensitivity check addresses this choice.

Christoffersen and Diebold \cite{christoffersen2006} distinguish conditional-mean, sign, and volatility forecastability and show how sign dependence can arise through volatility dynamics. This supports evaluating direction separately from return magnitude; it does not validate our particular council. Scale-aware evaluation also has a forecasting literature \cite{hyndman2006}. Our $E/V$ diagnostic uses historical volatility, not the naive-error denominator of mean absolute scaled error, and should not be called MASE.

\section{Study setting and cohort}
\subsection{A fixed historical council}
The frozen evaluation archive has an outcome cutoff of 8 October 2026. Its individual-forecast extraction covers Batches 3--7. We use completed native first-quarter forecasts from Batches 3--6, rather than daily refreshes, partially completed paths, or longer horizons. Batch 7, issued in September, is excluded because its native first-quarter targets had not matured. Batches 1--2 are outside this comparable ten-configuration extraction, not additional matched cohorts rejected on forecast performance; this study does not quantify their outcome attrition.

The eligible set contains ten configurations representing seven frameworks (Table~\ref{tab:frameworks}). These are platform persona labels for configured AI research perspectives, not forecasts made or endorsed by the named individuals. Four frameworks have a researcher mode; the other three have researcher and thinker modes. We first average modes within each framework, then average the seven framework forecasts. This gives each framework equal weight and prevents those with two modes from receiving twice the influence.

\begin{table}[H]
\centering\small
\begin{tabular}{llr}
\toprule
Framework label & Historical modes retained & Configurations\\
\midrule
Paradigm shift & Researcher & 1\\
Contrarian forensic & Researcher & 1\\
Power audit & Researcher & 1\\
Forensic deduction & Researcher & 1\\
Economic machine & Researcher; thinker & 2\\
Civilizational compute & Researcher; thinker & 2\\
Value ownership & Researcher; thinker & 2\\
\midrule
Total & Seven framework perspectives & 10\\
\bottomrule
\end{tabular}
\caption{Fixed historical council, with neutral framework names. Appendix~\ref{app:frameworks} maps archive labels and explains the perspectives. All configurations use archived Gemini versions; framework diversity is not provider diversity or statistical independence.}
\label{tab:frameworks}
\end{table}

The model-version identifiers change between the earlier and later pairs of batches; they are recorded in the companion data. Thus, ``fixed council'' means a fixed set of framework--mode configurations, not a constant model binary. Provider, prompt, and model-version effects are not separately identified. This study should not be read as a benchmark of the newest models or of all multi-agent architectures.

Of 1,242 candidate asset--batch date groups, 15 lack the complete ten-configuration set. The remaining 1,227 matched outcomes cover 376 assets. Requiring an asset in all four batches yields 233 assets and 932 outcomes. One asset lacks valid historical-risk coverage across all four batches, leaving the primary balanced panel of \textbf{232 assets and 928 outcomes}. The panel includes 206 equities, 14 crypto assets, five commodities, five foreign-exchange assets, and two funds. Each contributes four outcomes. Selection is based on completeness, not observed forecast accuracy, but requiring future coverage still introduces a retrospective survivorship constraint.

\begin{figure}[H]
\centering\includegraphics[width=\linewidth]{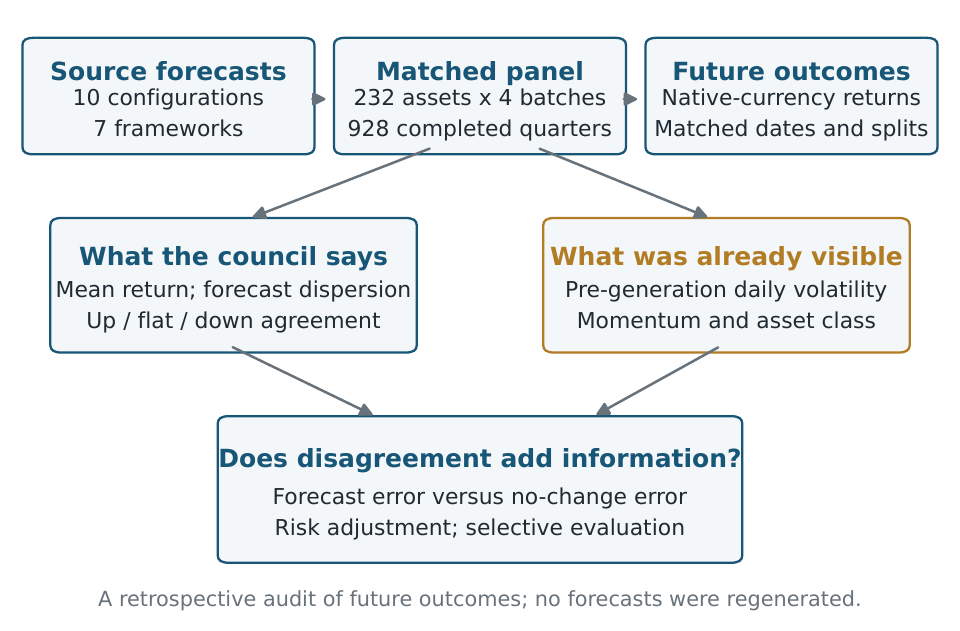}
\caption{Study design. Seven framework-level forecasts are compared with later outcomes and with risk measures available from prices dated before generation. The analysis distinguishes raw error, error relative to a baseline, and selective evaluation.}
\label{fig:design}
\end{figure}

\subsection{The information clock matters}
\begin{figure}[htbp]
\centering\includegraphics[width=\linewidth]{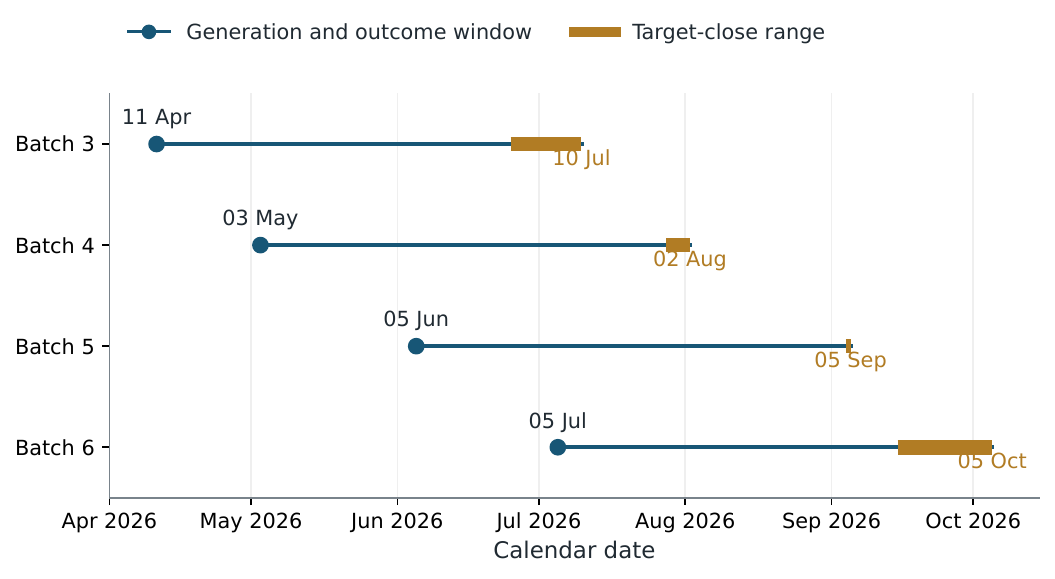}
\caption{Overlapping first-quarter outcome windows. Blue circles mark generation dates; gold segments mark the range of target-close observations across assets. Most preceding-quarter labels were unavailable at the next generation. The figure explains why adjusted models are treated as retrospective diagnostics.}
\label{fig:timing}
\end{figure}

Figure~\ref{fig:timing} shows the forecast and outcome windows. Much of Batch 3 had not matured when Batch 6 was generated, and the following batches had still less completed history available. Within the primary panel, only one earlier first-quarter outcome had closed before the Batch 6 issue date; none had closed before the Batch 4 or Batch 5 issue dates. Fitting an error predictor to all four batches and presenting it as a policy that could have been trained at each issuance would therefore be misleading.

Our retention rules require no fitted outcome labels. They rank assets within a batch using council dispersion, historical volatility, or their ratio. The later regression is explicitly a \emph{retrospective association analysis}, not a prospectively trained confidence model. Rules, cutoffs, and analyses were developed after an initial exploratory inspection; this study was not preregistered.

Forecast evaluation starts at the archive's first saved close after generation, rather than an assumed executable price at the exact generation time. Predicted target values and realized target prices are expressed relative to that common entry close. Consequently, these rebased scores become available at the entry close, not necessarily at the generation instant. The price-history risk measures use dates strictly before generation. This construction aligns forecasts with a common evaluable return, without implying that a live trade was placed.

\FloatBarrier

\section{Measures and analysis}
\subsection{Agreement, error, and the baseline}
For asset $a$ in batch $b$, let $p_{abj}$ be the predicted percentage return of framework $j$, after averaging its retained modes, and let $y_{ab}$ be the realized percentage return. With $J=7$, the council return and numerical disagreement are
\begin{equation}
\bar p_{ab}=\frac{1}{J}\sum_{j=1}^{J}p_{abj},\qquad
D_{ab}=\sqrt{\frac{1}{J}\sum_{j=1}^{J}(p_{abj}-\bar p_{ab})^2}.
\end{equation}
The population standard deviation describes the spread of this specified council. It is not a predictive standard error, an estimate based on seven independent draws, or a probability interval for the future return.

Absolute return error, no-change error, and their difference are
\begin{equation}
E_{ab}=|\bar p_{ab}-y_{ab}|,\qquad E^0_{ab}=|y_{ab}|,\qquad
\Delta E_{ab}=E_{ab}-E^0_{ab}.
\end{equation}
All are measured in percentage points (pp). Positive $\Delta E$ means the council does worse than predicting a zero return; negative values mean it does better. Subtracting this simple baseline does not remove every source of forecast difficulty. It does show whether a raw error pattern also occurs without using the advisors' predicted return.

Directional agreement is measured separately. A return above $+0.5$ pp is up, below $-0.5$ pp is down, and otherwise flat. Let $s(p)$ denote this classification. Then
\begin{equation}
A_{ab}=\max_{c\in\{-1,0,1\}}\frac{1}{J}\sum_{j=1}^{J}\mathbf{1}\{s(p_{abj})=c\}.
\end{equation}
High directional agreement means $A\geq6/7$. We evaluate two predictions separately: the sign of the \emph{council mean} and the \emph{modal framework direction}. For modal ties we fall back to the mean sign, and also report accuracy after excluding ties. Mean and modal direction need not coincide. We compare both with an always-up directional baseline; no change is the separate return-error baseline. V2 adds the modal sensitivity to the original mean-sign analysis rather than selecting a better-performing rule retrospectively.

\subsection{Visible asset risk}
Historical volatility is computed from the most recent 60 observed daily log returns before generation, requiring at least 45 observations. Recorded active splits are incorporated into the historical returns. A conservative check excludes histories with an unsupported large split adjustment; 20- and 90-observation windows provide sensitivity comparisons.

To place volatility on the approximate horizon scale, define
\begin{equation}
V_{ab}=100\,\hat\sigma_{ab}\sqrt{h_{ab}\,n_a/365},
\end{equation}
where $h$ is the number of calendar days from entry to target, and $n_a$ is 365 for crypto assets and 252 otherwise. This square-root scaling is a conventional descriptive risk measure, not a claim that returns are Gaussian or independent. Historical momentum is the cumulative return over the same 60-observation window. All outcomes are native-currency price returns; they do not include dividend reinvestment or a common-currency portfolio conversion.

\subsection{Comparisons and uncertainty}
We assign disagreement quartiles \emph{within each batch}, so a change in the overall forecast scale cannot alone create the pooled gradient. Each primary quartile contains 58 assets per batch, or 232 outcomes. We also form five within-batch volatility groups and compare lower and higher disagreement within each group.

The adjusted model relates $\log(1+E)$ to historical volatility, absolute forecast magnitude, absolute historical momentum, and disagreement, with batch and asset-class indicators. The logarithm reduces the influence of very large errors and defines an association on a transformed scale; it is not equivalent to modeling MAE. Continuous predictors are transformed and standardized; exact definitions appear in Appendix~\ref{app:reproduce}. A second model uses $\Delta E$ as the response; v2 also fits untransformed $E$ as a sensitivity. These controls test conditional associations, not a causal effect, complete risk removal, or absence of a small useful signal.

For selective evaluation, we rank $D$, $V$, and $D/V$ within each batch. V2 expands the original 25/50/75/100\% comparisons to 5\% increments and adds a residual-spread diagnostic: residuals from within-batch OLS of $\log(1+D)$ on an intercept and $\log V$. This uses no outcome labels but fits the observed cross-section, so it is not a validated deployment policy. We also report mean $E/V$ and the pooled skill ratio $1-\sum E/\sum E^0$. The latter avoids unstable per-record ratios when $E^0$ approaches zero. These descriptive coverage comparisons contain no position sizing, execution, transaction costs, or utility objective.

We report pointwise 95\% percentile intervals from 5,000 asset-cluster bootstrap draws, resampling entire assets with their four outcomes together. Regression intervals use 2,000 draws. Paired contrasts use the same draws. The original fixed-membership intervals describe the selected archived groups. A second bootstrap recomputes within-batch ranks and retained membership in every sampled asset population, addressing selection-rule variability under asset resampling. Both condition on these four observed periods; neither represents arbitrary common market shocks or new regimes. We therefore report period-specific and leave-one-batch-out point sensitivities, rather than treating four overlapping batches as sufficient for reliable two-way-cluster inference. Intervals are pointwise, exploratory, and not multiplicity-adjusted. A further sensitivity resamples 16 whole equity-sector/non-equity-class blocks with all their assets and periods. This grouping preserves within-block dependence, holds archived ranks fixed, and uses frozen metadata; it cannot capture a shock shared by every group or provide future-regime inference.

\section{Results}
\subsection{Establishing the council's overall skill}
Across the full panel, council MAE is 15.15 pp versus 14.99 pp for no change. The paired difference is +0.16 pp [$-0.23$, 0.56]. Council mean-sign accuracy is 48.38\%, compared with 50.75\% for always up. These comparisons provide no clear evidence of an overall advantage under the chosen losses; they do not establish equivalence or evaluate qualitative research. Median aggregation has MAE 15.13 pp. Averaging the seven separate framework MAEs gives 15.68 pp, and individual framework MAEs range from 15.08 to 17.01 pp. Aggregation improves on the average member descriptively, but that is a different comparison from improving on a simple baseline.

Forecast magnitude matters for interpreting excess error. The median absolute council forecast is 3.16 pp (mean 4.20), while realized returns range from $-52.10$ to +205.78 pp. The reverse triangle inequality gives $|E-E^0|\leq|\bar p|$: forecasts close to zero necessarily produce losses close to no change. This does not force a zero difference or invalidate the benchmark, but limits how much incremental loss separation these forecasts can exhibit. Forecast--outcome Pearson correlation is 0.027 in this panel. Appendix~\ref{app:diagnostics} reports distributions and framework results.

As an additional unfitted comparator, repeating the preceding 60-observation cumulative return yields MAE 25.47 pp. Its lookback differs from some target lengths and it is neither an optimized momentum model nor a comprehensive finance benchmark. The council's lower loss against this rule cannot substitute for its unresolved difference against no change.

\subsection{The apparent confidence signal mirrors a simple baseline}
\begin{figure}[htbp]
\centering\includegraphics[width=\linewidth]{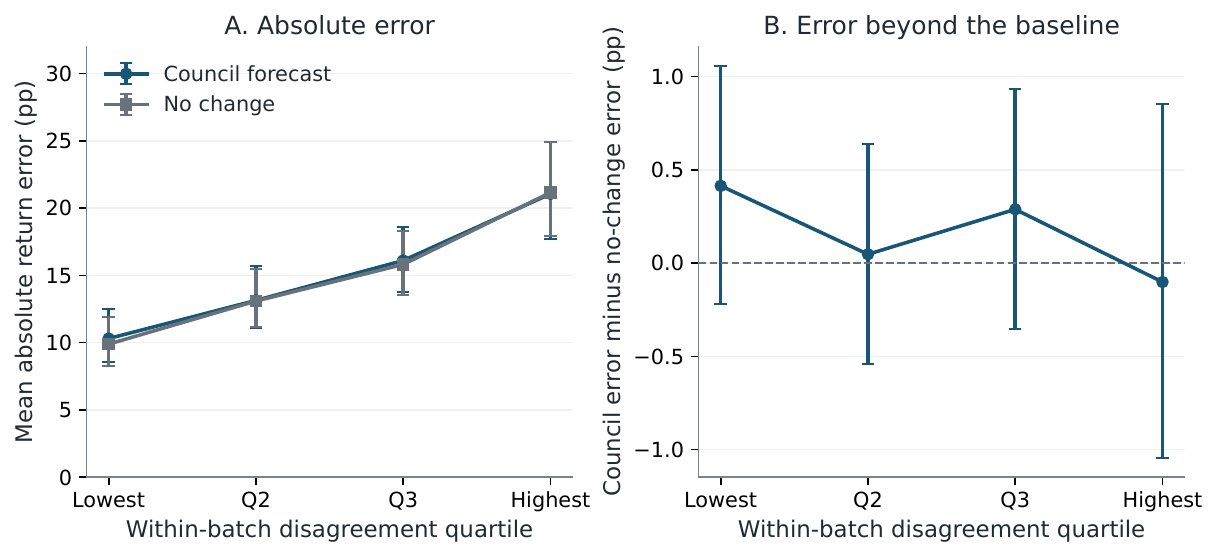}
\caption{The raw disagreement--error gradient is also present for no change. A: absolute errors for the council and baseline. B: their paired difference within each quartile. Bars are pointwise asset-cluster 95\% intervals.}
\label{fig:quartiles}
\end{figure}

Numerical disagreement strongly separates later absolute error. Mean error rises from 10.31 pp in the lowest-disagreement quartile to 21.05 pp in the highest (Figure~\ref{fig:quartiles}). The high-minus-low gap is 10.74 pp, with a 95\% interval of [7.11, 14.78]. Viewed alone, this looks like a useful confidence signal.

The no-change baseline, however, rises almost identically: 9.89 pp to 21.15 pp. Its high-minus-low gap is 11.26 pp [7.82, 15.10]. The corresponding gap in error \emph{beyond the baseline} is $-0.52$ pp [$-1.74$, 0.69]. Thus, the raw gradient does not establish that low disagreement identifies greater council skill. Much of the separation also describes how far prices subsequently move from their starting values.

\begin{table}[H]
\centering\small
\begin{tabular}{lrrrr}
\toprule
Disagreement quartile & Council MAE & No-change MAE & Difference & Mean $V$\\
\midrule
Lowest & 10.31 & 9.89 & +0.41 & 14.31\\
Second & 13.14 & 13.09 & +0.05 & 16.91\\
Third & 16.10 & 15.81 & +0.29 & 21.21\\
Highest & 21.05 & 21.15 & $-0.10$ & 30.03\\
\bottomrule
\end{tabular}
\caption{Within-batch disagreement quartiles, pooled across four batches; 232 outcomes per row. Return errors and horizon-scaled volatility $V$ are in pp. Rounded entries may not subtract exactly.}
\label{tab:quartiles}
\end{table}

The recomputed-rank bootstrap gives an excess-error gap interval of [$-1.70$, 0.72] pp, compared with [$-1.74$, 0.69] under fixed membership. Accounting for rank selection under asset resampling does not resolve the excess-error gradient. These intervals still condition on the same four periods.

\subsection{Disagreement tracks risk visible before generation}
\begin{figure}[htbp]
\centering\includegraphics[width=\linewidth]{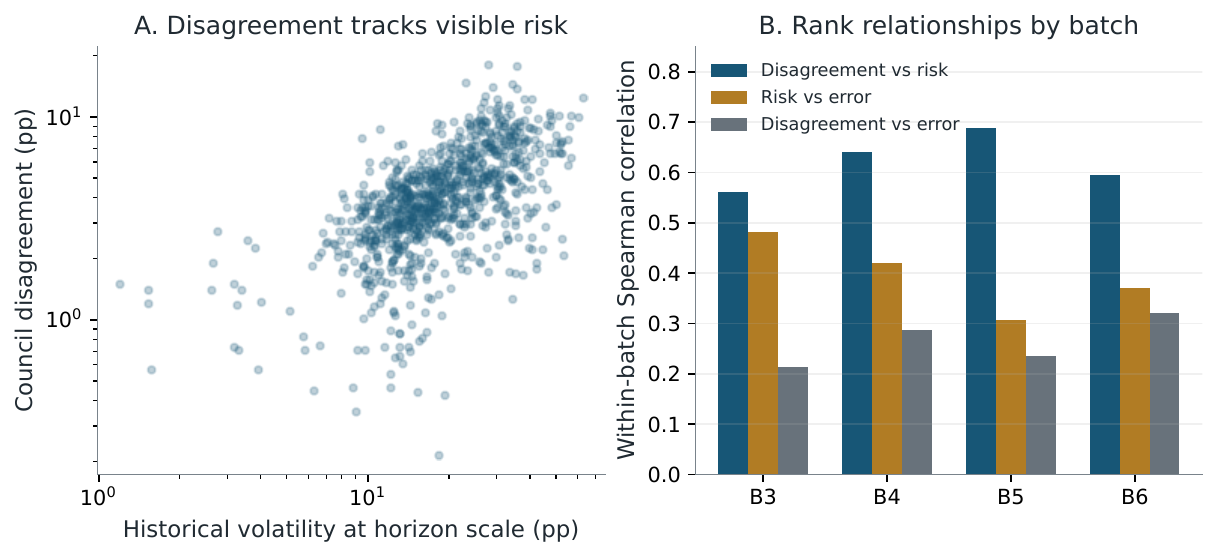}
\caption{Visible asset risk and council dispersion. A: 928 outcomes plotted on logarithmic axes. B: within-batch rank correlations. The council's spread is closely related to historical volatility, which is more strongly related to absolute forecast error than spread is in each batch.}
\label{fig:risk}
\end{figure}

The raw gradient has a plausible, observable explanation. Disagreement and historical volatility have within-batch Spearman correlations between 0.56 and 0.69 (Figure~\ref{fig:risk}). Historical volatility also correlates with subsequent council error, more strongly than disagreement does in each of the four batches. Advisors tend to spread their forecasts more widely on assets whose recent price behavior already signals greater variability.

In the adjusted model, the standardized log-disagreement coefficient for $\log(1+E)$ is 0.006 [$-0.072$, 0.079], while the historical-volatility coefficient is 0.358 [0.259, 0.465]. The disagreement coefficient in the excess-error model is $-0.08$ pp [$-0.63$, 0.42]. These estimates do not resolve an additional association after the specified controls; they are not an equivalence test.

The coarse, less model-dependent comparison gives a similar result. Within volatility groups, the higher-minus-lower disagreement mean-error contrast is $-1.02$ pp [$-3.99$, 1.79], spanning both directions. Their excess-error contrast is $-0.31$ pp [$-1.04$, 0.41]. Five risk bins permit comparing spread halves at broadly similar risk while retaining usable cell sizes; four disagreement quartiles summarize the unconditional gradient. Neither scheme eliminates all continuous-risk differences.

\begin{figure}[htbp]
\centering\includegraphics[width=\linewidth]{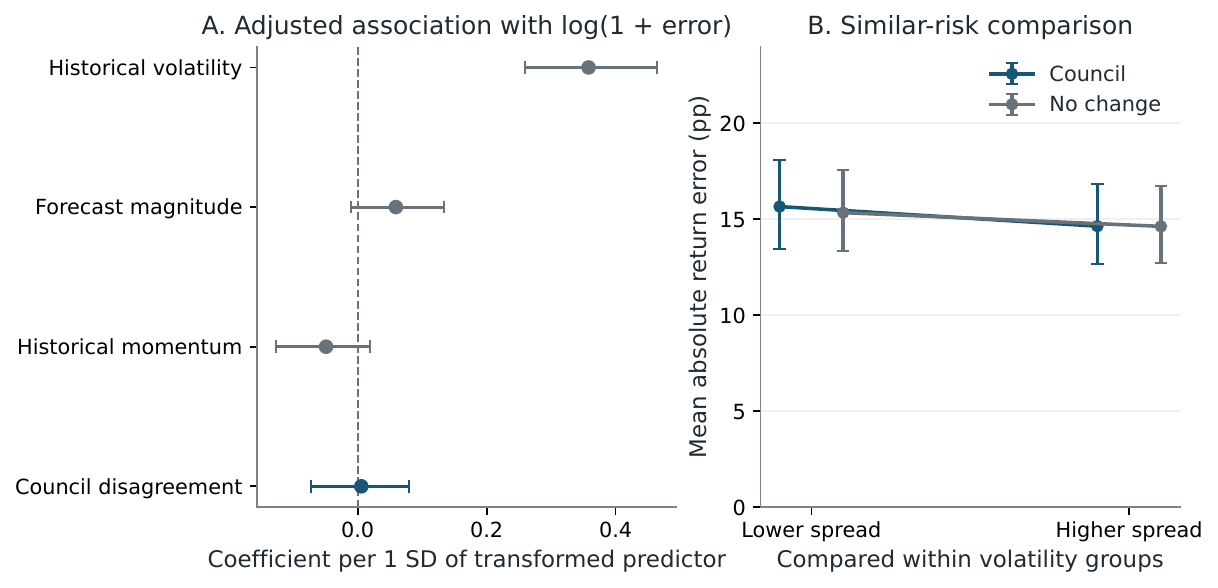}
\caption{Adjusted and similar-risk comparisons. A: coefficients for standardized transformed predictors in the model of $\log(1+E)$, controlling for batch and asset class. B: lower and higher disagreement within five volatility groups per batch. Bars show pointwise asset-cluster 95\% intervals. These are retrospective diagnostics.}
\label{fig:adjusted}
\end{figure}
\FloatBarrier

\subsection{Historical volatility selects lower-error forecasts}
\begin{figure}[htbp]
\centering\includegraphics[width=\linewidth]{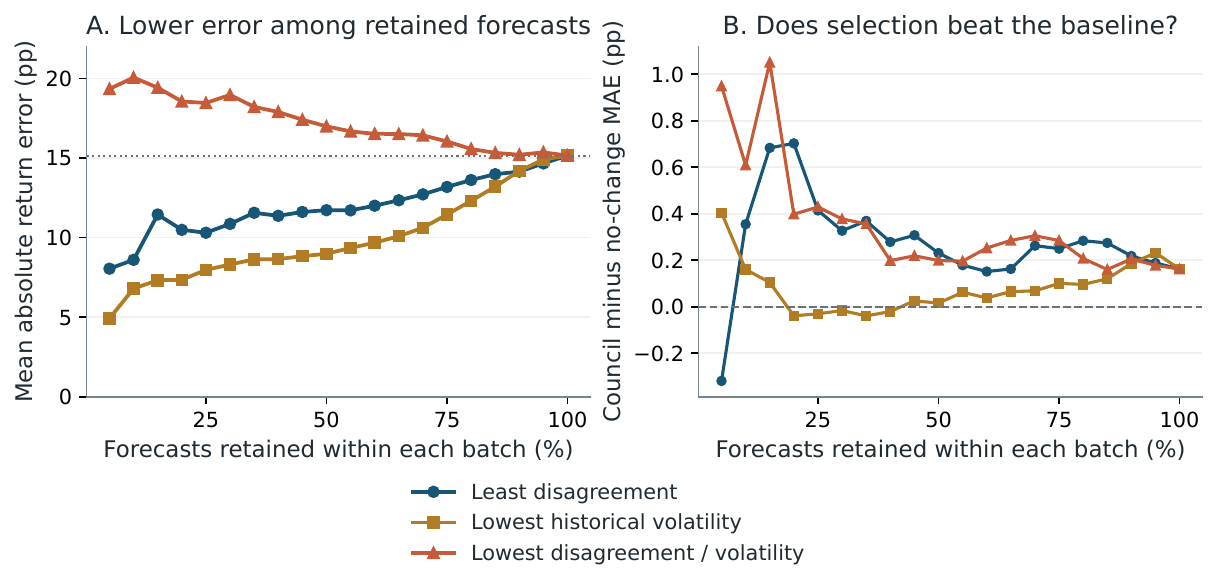}
\caption{Selective evaluation at equal within-batch coverage. A: absolute error among retained forecasts. B: paired error relative to no change. Points are observed estimates at 5\% coverage increments; these curves have no uncertainty bands. Paired selector intervals are reported in the text. At 100\% all rules retain the same panel. The rules require no fitted outcome labels; their evaluation is retrospective.}
\label{fig:coverage}
\end{figure}

Selecting the least-disagreeing quarter of forecasts reduces MAE from 15.15 pp to 10.31 pp. Selecting the lowest-volatility quarter reduces it further, to 7.97 pp (Figure~\ref{fig:coverage}). The paired difference between the volatility and disagreement selectors is $-2.34$ pp [$-4.26$, $-0.80$] under fixed membership; recomputing selection gives [$-4.34$, $-0.98$]. Both rules retain 58 assets per batch, but need not retain the same assets.

The pooled selector advantage is period-sensitive. Leaving out Batch 3 reduces the volatility-minus-disagreement MAE difference to $-0.02$ pp; omitting Batches 4, 5, or 6 gives $-2.66$, $-3.18$, or $-3.49$ pp. Resampling 16 larger cross-asset sector/class blocks gives an interval of [$-5.68$, $-0.04$] pp under fixed archived selection. This sensitivity preserves more within-group co-movement, but few unequal groups and unchanged market periods limit its interpretation. The pooled advantage should not be presented as stable prospective performance.

This improvement is about selecting easier cases, not demonstrating greater forecasting skill. On the lowest-volatility quarter, no change has MAE 8.00 pp; the council-minus-baseline difference is $-0.03$ pp [$-0.43$, 0.37]. On the least-disagreeing quarter, the council's difference is +0.41 pp. A selection rule can substantially reduce error while leaving the council's advantage over the baseline unresolved.

Dividing disagreement by volatility does not automatically produce a better confidence signal. Keeping the lowest $D/V$ quarter yields MAE 18.45 pp. A large denominator can make a volatile asset appear to have small relative disagreement. This failed shortcut is useful: controlling for risk in an analysis and constructing a reliable selection score are different tasks.

\subsection{Directional agreement depends on the aggregation rule}
High agreement occurs in 436 outcomes; the other 492 have fewer than six frameworks sharing a direction. For the sign of the arithmetic mean, accuracy is 53.21\% versus 44.11\%, a +9.11-point contrast [3.03, 15.22]. But agreement itself counts categorical votes, while the arithmetic mean also gives large numerical forecasts more influence. These are distinct ways to turn the same council into a prediction.

\begin{figure}[htbp]
\centering\includegraphics[width=\linewidth]{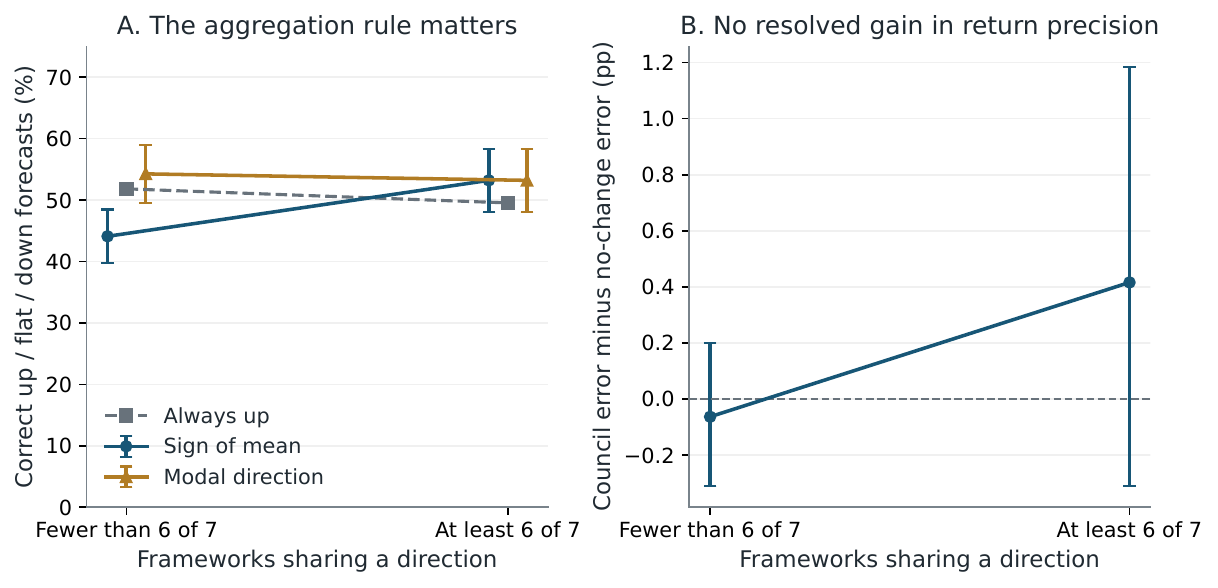}
\caption{Two aggregation rules, two directional conclusions. A: mean-sign and modal-direction accuracy for the same agreement groups; modal ties fall back to mean sign. B: paired council-minus-no-change MAE, rather than visually inferred differences between raw-loss markers. Bars are pointwise asset-cluster 95\% intervals, conditional on the observed periods.}
\label{fig:direction}
\end{figure}

Scoring the modal direction leaves high-agreement accuracy at 53.21\% but raises lower-agreement accuracy to 54.27\% (Figure~\ref{fig:direction}). The contrast becomes $-1.06$ points [$-7.53$, 5.30]. In 108 of 901 untied cases the mean and modal predictions differ; all 108 are in the lower-agreement group. There are 27 modal ties. Excluding those ties gives lower-agreement modal accuracy of 55.91\% on 465 outcomes. No mean/modal mismatch occurs among the 436 high-agreement outcomes. Thus, the original positive association describes \emph{mean-sign accuracy}, not a general benefit of agreement for majority-vote direction.

Return magnitude supplies another distinct estimand. Council/no-change MAEs are 16.84/16.42 pp under high agreement and 13.65/13.71 under lower agreement. The high-minus-lower \emph{raw} error difference is +3.19 pp, but the difference in \emph{excess} error is only +0.48 pp [$-0.26$, 1.23]. Larger raw error therefore does not establish worse baseline-relative skill, just as the directional contrast does not establish better numerical precision.

\begin{figure}[htbp]
\centering\includegraphics[width=\linewidth]{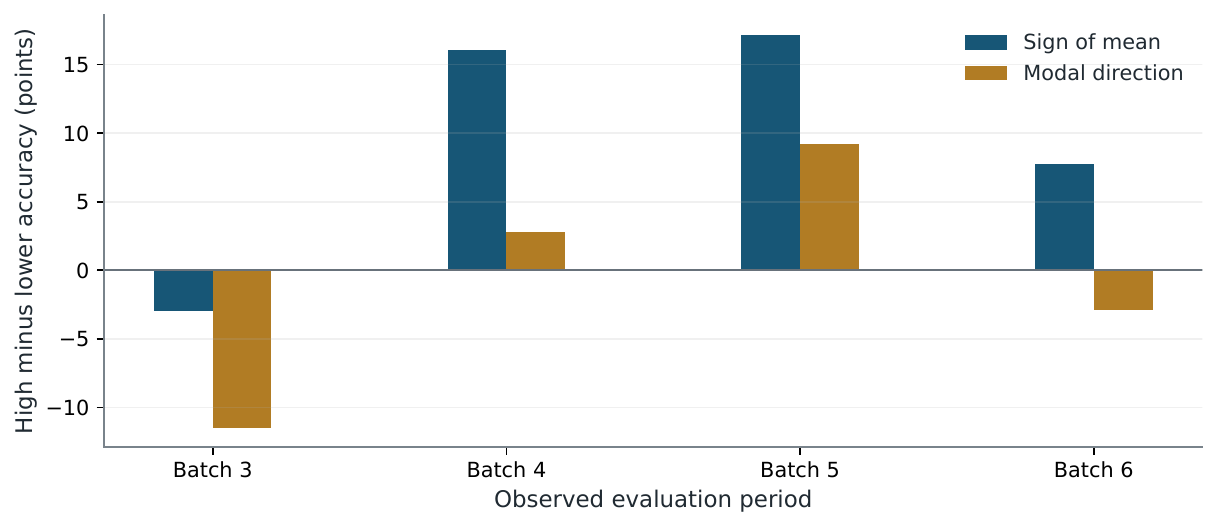}
\caption{Directional association varies across periods and aggregation rules. Each bar is high-minus-lower agreement accuracy within one batch (232 outcomes). These observed-period contrasts have no future-regime uncertainty interpretation.}
\label{fig:periods}
\end{figure}

The positive mean-sign contrast is heterogeneous: $-2.99$, +16.09, +17.18, and +7.72 points in Batches 3--6. Modal contrasts are $-11.46$, +2.76, +9.24, and $-2.90$ points (Figure~\ref{fig:periods}). Leaving out one batch gives mean-sign contrasts from +6.56 to +13.55 points, but modal contrasts from $-4.37$ to +2.86. Coarse comparisons within five risk groups per batch give population-size-weighted contrasts of +8.94 points for mean sign and $-0.72$ for modal direction; signed-momentum groups give +9.68 and $-0.04$. These are descriptive strata averages, not causal adjustment or new significance tests. Equity-only and expanded-cohort results, confusion matrices, all attainable agreement levels, and a complete sign-band/cutoff grid are released. The grid is a sensitivity analysis, not a search for an optimal confidence rule.

\FloatBarrier
\subsection{Robustness and variation across batches}
\begin{figure}[htbp]
\centering\includegraphics[width=\linewidth]{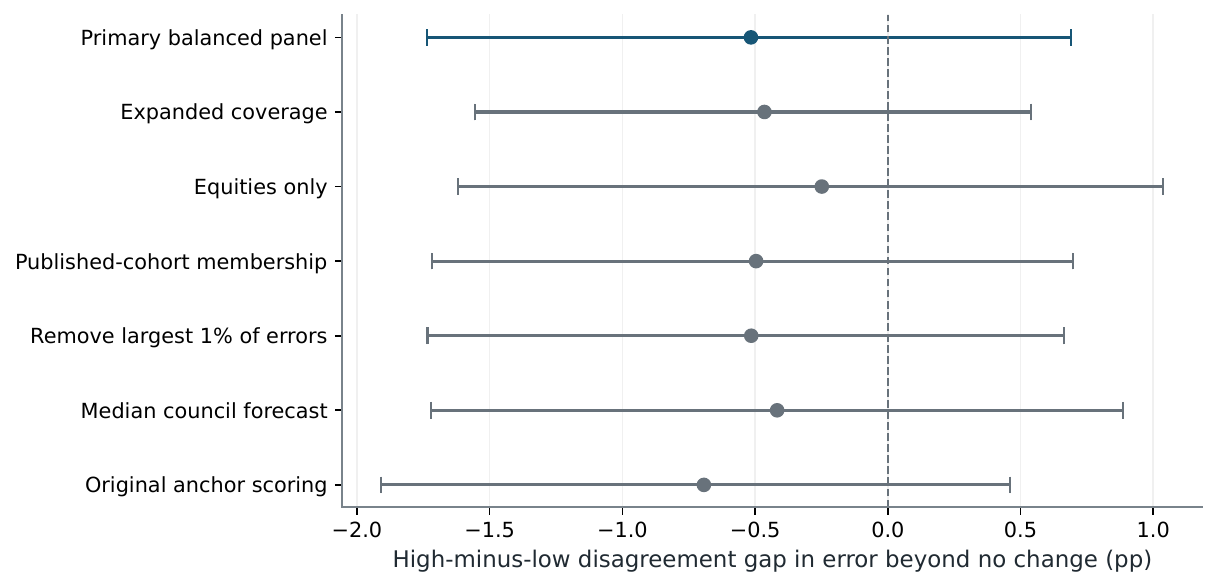}
\caption{Sensitivity of the high-minus-low disagreement contrast in excess error. Negative values indicate a smaller council-minus-baseline error in the highest-disagreement quartile. Pointwise asset-cluster 95\% intervals span zero across the listed specifications.}
\label{fig:robustness}
\end{figure}

The central excess-error contrast remains unresolved when we expand coverage to 1,221 outcomes, restrict to equities, require original published-cohort membership, remove the largest 1\% of errors, replace the council mean with its median, or score against the archived original anchor (Figure~\ref{fig:robustness}). Every corresponding interval spans zero. The reconstruction audit identifies two Batch 3 commodity records whose forecast rebasing factors differ from the historical outcome-anchor ratios. Restricting to the other 926 records gives an entry-based gap of $-0.51$ pp [$-1.73$, 0.70] and an anchor-based gap of $-0.66$ pp [$-1.86$, 0.49]. The primary entry-based scoring is preserved, while compatible-anchor scoring is reported separately. This distinction prevents silently treating all recorded forecast anchors as identical to daily historical closes. Alternative spread measures also give unresolved high-minus-low excess-error gaps: +0.11 pp [$-0.92$, 1.17] for the interquartile range and $-0.42$ pp [$-1.46$, 0.61] for median absolute deviation.

The lowest-volatility quarter remains a low-error subset with 20- and 90-observation windows: MAE is 7.60 and 7.50 pp, respectively. Their council-minus-no-change differences are small, with intervals spanning zero. Root mean squared errors rise across disagreement quartiles for both the council and baseline as well (Appendix~\ref{app:tables}). Thus, the visible raw gradient is not peculiar to a single absolute-error calculation.

The pooled council--baseline comparison also varies by batch. Batch 3 has council MAE 18.88 pp versus 16.97 pp for no change; the other three batches have small mean advantages for the council (Table~\ref{tab:batches}). This variation cautions against interpreting the pooled near-zero difference as a stable characteristic of the platform. The available record contains only four overlapping evaluation periods.

\FloatBarrier

\subsection{Scale and framework dependence}
\begin{figure}[htbp]
\centering\includegraphics[width=\linewidth]{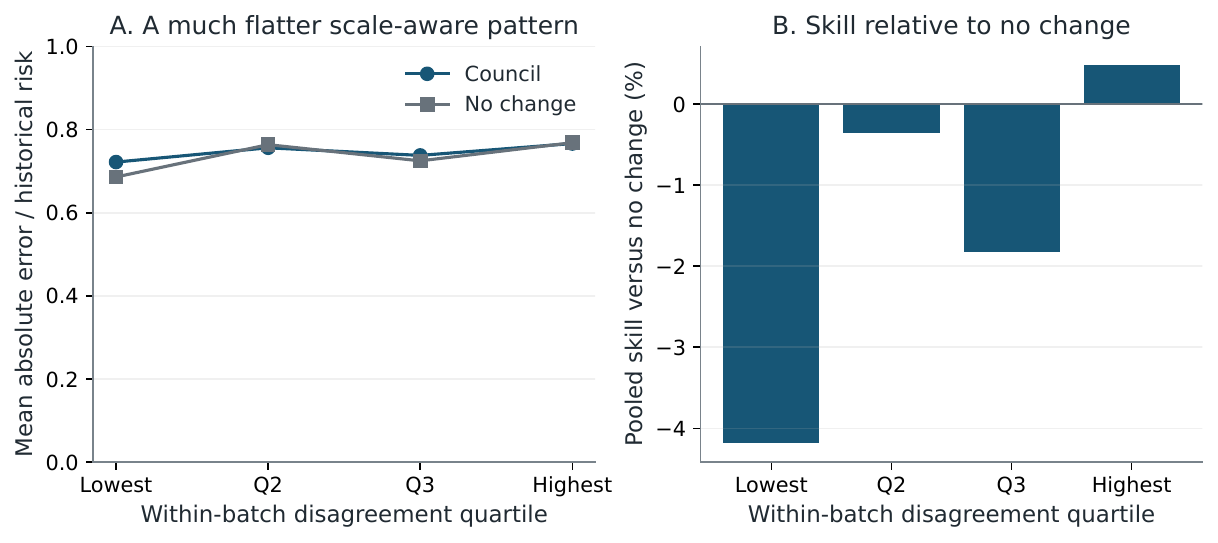}
\caption{Scale-aware diagnostics for the same quartiles. A: mean per-record error divided by historical horizon-scale volatility. B: pooled council skill relative to no change, $1-\sum E/\sum E^0$; positive favors the council. Points and bars are descriptive estimates without uncertainty bands.}
\label{fig:scaled}
\end{figure}

Scaling each loss by its pre-generation risk measure makes the disagreement-quartile pattern much flatter: mean $E/V$ is 0.722, 0.756, 0.738, and 0.767. The pooled baseline skill ratios are $-4.19$, $-0.36$, $-1.82$, and +0.48\%, respectively (Figure~\ref{fig:scaled}). These dimensionless diagnostics complement rather than replace the common pp loss. Dividing spread by risk, or residualizing it against risk, does not itself provide a calibrated uncertainty measure; their full retention curves are included in the companion.

\begin{figure}[htbp]
\centering\includegraphics[width=.88\linewidth]{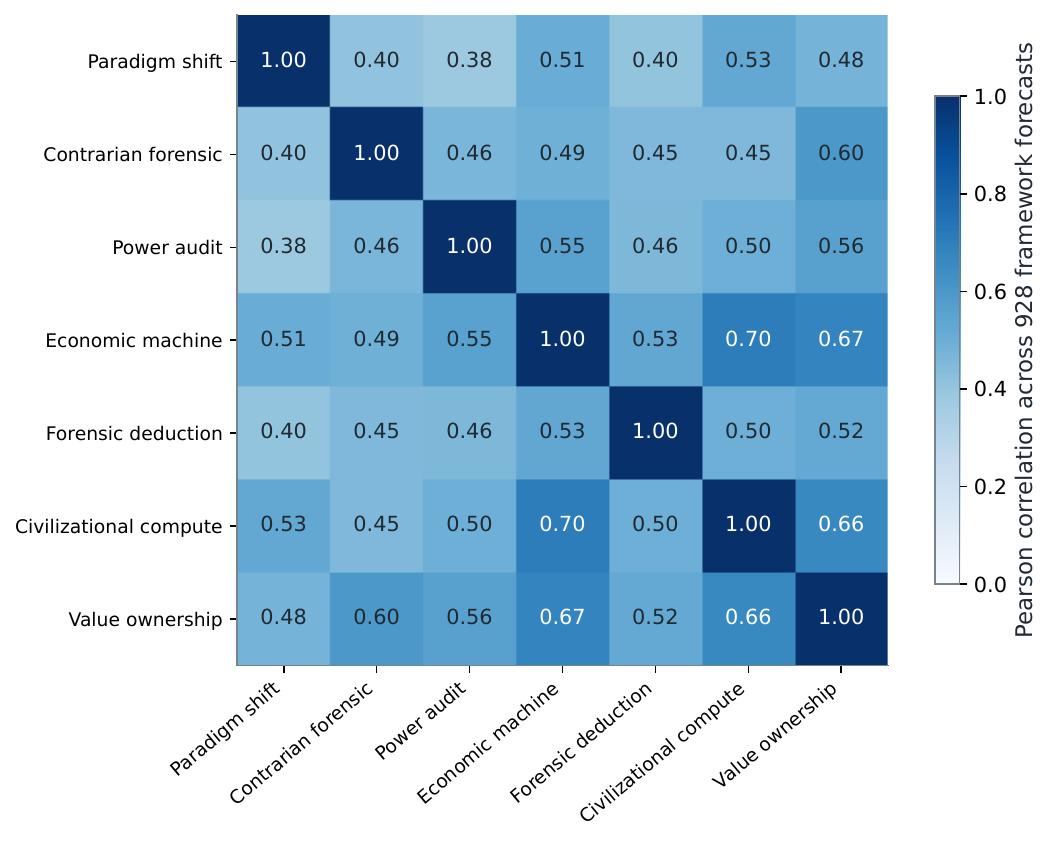}
\caption{Dependence among framework return forecasts across 928 outcomes. Pearson correlations pool four batches; the companion also supplies matrices within each batch. Neutral names map to historical archive labels in Appendix~\ref{app:frameworks}. Correlation is not a measure of causal independence or model quality.}
\label{fig:dependence}
\end{figure}

Pooled pairwise framework forecast correlations range from 0.378 to 0.700 (Figure~\ref{fig:dependence}). They reflect shared assets and conditions as well as potential shared model behavior; they do not isolate independent reasoning or error correlations conditional on context. Within the three two-mode frameworks, mode correlations are 0.505, 0.458, and 0.635. The variance of the mode average is 0.751, 0.729, and 0.817 times the average mode variance. Averaging modes therefore does not empirically halve forecast variance, nor does forecast variance identify a decomposition into signal and noise.

\FloatBarrier

\section{What an agreement score should communicate}
The results suggest separating three concepts in an AI forecasting system. \emph{Asset risk} concerns how variable the outcome may be. \emph{Directional concordance} concerns whether the advisors share an up, flat, or down view. \emph{Return precision} concerns how close their numerical prediction is to the eventual return. A score that blends these concepts can appear predictive while obscuring what it actually predicts.

For this council, historical volatility is a strong first reference for forecast difficulty. Council dispersion adds no clearly resolved relationship with excess error under the tested comparisons. Directional agreement has a positive association with mean-sign accuracy in the pooled sample, but not with modal-direction accuracy. The conclusion is therefore more specific than ``agreement is useless'': different forms of agreement carry different information, and none should be advertised as universal confidence without further validation.

A sensible next test would freeze the feature definitions and thresholds before a new batch, register the available training labels, and evaluate direction and return losses separately after the native targets mature. If an error predictor is learned, validation must respect both the issue date and the maturity date of every label. A model trained on future-resolved labels cannot be presented as a real-time confidence layer merely because the forecasts themselves were made earlier.

This archive does not contain the consistent predictive distributions needed to establish probabilistic calibration. A seven-framework spread cannot be labeled an 80\% or 95\% confidence interval without an additional, validated mapping to outcome coverage. Similarly, being more accurate on a retained subset does not establish an economic advantage. Such claims require a defined decision rule and appropriate costs, baselines, and out-of-time testing.

\section{Limitations}
\paragraph{One system and few periods.} The council shares historical Gemini model versions, and the four forecast periods overlap. Framework diversity is neither independence nor diversity across providers. Asset-cluster intervals preserve repeated assets but cannot eliminate dependence induced by common market shocks. Generalization to other models, markets, horizons, or regimes remains untested.

\paragraph{Retrospective coverage and data vintage.} Assets were curated for a production platform and then required to have complete forecasts and history. They are not a random sample of the investable universe. The evaluation uses a frozen October archive, not an immutable contemporaneous market-data vintage for every past day. Revised prices, corporate actions, and reconstructed context can differ from what was available during the original run. The analysis respects dates in the archive but cannot independently certify the full historical information set of every prompt.

\paragraph{Scoring and operational choices.} Native-currency price returns exclude dividend reinvestment, currency conversion, and execution costs. Evaluation is rebased to a saved post-generation close, rather than an assumed trade at the generation instant. The council mean, volatility scaling, sign threshold, and retention fractions are choices; the reported sensitivities address some, not all, alternatives. Qualitative report quality and synthesized portfolio recommendations are outside the scope.

\paragraph{Exploration and unresolved effects.} The analyses follow an exploratory audit rather than a preregistration. Multiple secondary comparisons are not multiplicity-adjusted. Intervals overlapping zero do not demonstrate equivalence or rule out small useful effects. In particular, the directional-agreement result merits prospective confirmation before it is used as a confidence rule.

\begin{samepage}
\section{Conclusion}
In this historical AI council, wider numerical disagreement predicts larger return errors, but nearly the same pattern occurs for a no-change forecast and is strongly associated with visible asset volatility. A simple volatility rule selects lower-error forecasts more effectively than raw disagreement. The positive directional association is specific to the sign of the arithmetic mean; it disappears under modal voting. Overall council MAE remains close to no change, and neither directional concordance nor low numerical spread establishes incremental forecasting skill. The practical lesson is to evaluate agreement against the question it is meant to answer. Before calling consensus a confidence signal, establish whether it measures ordinary outcome risk, shared directional judgment, or an actual improvement in forecast reliability.

\end{samepage}
\section*{Data, code, and author disclosure}
The \href{https://doi.org/10.5281/zenodo.23373782}{Zenodo research deposit} (DOI: \href{https://doi.org/10.5281/zenodo.23373782}{10.5281/zenodo.23373782}) and \href{https://github.com/TheFutureEdge/research-when-ai-advisors-agree}{public research repository} contain individual configuration forecasts, cohort metadata, aggregate results, deterministic analysis code, and vector figures. The frozen source-archive fingerprint and extraction rules are documented. The full market-derived outcome panel, raw market-history files, credentials, and operational metadata are excluded from the public release. Recomputing outcome-dependent results requires compatible market inputs obtained under appropriate terms or authorized access to the author-side panel. Historical values come from the iPulse AI market-history archive; no alternative source attribution is implied. Original manuscript, figures, forecast data, and research summaries use CC BY 4.0; code uses MIT. Third-party material retains its applicable terms.

The author is the founder of Future Edge Group FZE, which develops iPulse AI, and has a commercial interest in the platform studied. AI systems assisted with research operations, coding, audit, and manuscript preparation. Authorship and responsibility for the final submission remain with the named human author. The forecasts evaluated here come from historical Gemini configurations; OpenAI Codex assisted with the v2 analysis, code, checks, and manuscript preparation. These roles are distinct, and the revision generates no new forecasts.

\paragraph{Interpretation of small differences.} Failure to resolve an excess-error effect is not evidence of equivalence. For illustration, the recomputed-rank gap interval [$-1.70$, 0.72] pp extends beyond both $\pm0.5$ and $\pm1$ pp bounds. Although it lies within $\pm2$ pp, that bound was not chosen prospectively and the interval is conditional on observed periods. No practical equivalence claim follows. A future evaluation should predefine its loss, aggregation rule, meaningful effect margin, cohort, and market-period sampling plan; collect additional non-overlapping mature batches; and keep a final period untouched.

\clearpage
\appendix
\section{Reproduction and cohort details}\label{app:reproduce}
The outcome cutoff remains 8 October 2026. The original manuscript was dated 9 October; this revision is dated 14 October. No new outcome or forecast generation has been added. The input is identified by SHA-256\footnote{\texttt{f4f9510fe0ae2900f124b12e638f06eac073f0397c5646e6dfea62a278555be1}.}. Source extraction selects completed native first-quarter observations in Batches 3--6 with the eligible ten configurations. Matching requires one observation per configuration, identical entry and target dates, a common issue date, finite predictions, and realized returns agreeing to tolerance $10^{-6}$. No forecast is regenerated.

The expanded matched panel has 242, 307, 303, and 375 outcomes in Batches 3, 4, 5, and 6. The primary panel has 232 in each. The one balanced asset excluded for risk validity contributes four outcomes; its omission is not based on its forecast error. The author-side derived panel preserves exclusion flags and coverage variables for auditing; the public cohort file identifies matched records and primary membership.

Daily historical returns use positive finite closes before the issue date. Between consecutive observations, split ratios with recorded active ex-dates are applied. Histories are quarantined when a split adjustment produces an absolute daily log move above 0.4 while the unadjusted move is below 0.2. This heuristic conservatively avoids a possible double adjustment; it is not a general corporate-action validator. Momentum uses the last 60 available log returns. Historical volatility uses the sample standard deviation and at least 75\% of the requested observation window, with a minimum of 15 observations.

For the adjusted model, continuous variables are $\log V$, $\log(1+|\text{momentum}|)$, $\log(1+|\bar p|)$, and $\log(1+D)$. Each is standardized using the primary panel's mean and population standard deviation. Ordinary least squares includes batch and asset-class indicators; bootstrap refits preserve each asset's complete cluster. The random seed is 20261009. No fitted model is claimed to be deployable at historical issue dates.

A numerical tolerance of $10^{-10}$ pp at the $\pm0.5$ pp sign boundaries prevents floating-point classification artifacts.

For retained fractions, assets are sorted within each batch by score, breaking ties with the stable asset identifier; the retained count is rounded upward. The original bootstrap holds this membership fixed; the v2 selection-rule bootstrap rebuilds within-batch ranks after resampling whole assets. Fixed-membership and new-population selection targets are labeled separately. For the within-risk contrast, assets are sorted into five approximately equal-sized volatility bins per batch, then split into lower and higher disagreement halves within each bin. These finite-sample comparisons cannot eliminate all continuous-risk differences.

\section{Additional numerical results}\label{app:tables}
\begin{table}[H]
\centering\small
\begin{tabular}{lrrrrr}
\toprule
Batch & Council MAE & No-change MAE & Difference & Direction & Always up\\
\midrule
3 & 18.88 & 16.97 & +1.91 & 38.36\% & 52.16\%\\
4 & 13.70 & 14.21 & $-0.51$ & 50.43\% & 51.72\%\\
5 & 13.47 & 14.08 & $-0.62$ & 53.88\% & 59.91\%\\
6 & 14.54 & 14.68 & $-0.14$ & 50.86\% & 39.22\%\\
\bottomrule
\end{tabular}
\caption{Batch-level estimates in the primary panel, 232 outcomes per row. Return errors are in pp. Direction uses the three-category sign rule; always up is a separate directional baseline.}
\label{tab:batches}
\end{table}

\begin{table}[H]
\centering\small
\begin{tabular}{lrr}
\toprule
Disagreement quartile & Council RMSE (pp) & No-change RMSE (pp)\\
\midrule
Lowest & 17.69 & 16.55\\
Second & 21.01 & 20.07\\
Third & 23.39 & 22.79\\
Highest & 31.14 & 30.63\\
\bottomrule
\end{tabular}
\caption{Squared-error sensitivity. Both series exhibit increasing error across disagreement quartiles. This table supplies a second loss function, not a fitted probabilistic forecast evaluation.}
\end{table}

\begin{table}[H]
\centering\small
\begin{tabular}{lrrr}
\toprule
Volatility window & Retained outcomes & Council MAE (pp) & Excess MAE (pp)\\
\midrule
20 observations & 232 & 7.60 & +0.08 [$-0.31$, 0.45]\\
60 observations & 232 & 7.97 & $-0.03$ [$-0.43$, 0.37]\\
90 observations & 232 & 7.50 & +0.02 [$-0.36$, 0.41]\\
\bottomrule
\end{tabular}
\caption{Lowest-volatility quarter under alternative historical windows. Brackets are pointwise asset-cluster 95\% intervals for the paired error difference relative to no change.}
\end{table}

\FloatBarrier
The public CSVs permit checking forecast grouping and cohort membership. The v2 companion adds a documented market-input adapter, aggregate diagnostics, and a change log. Compatible independently sourced inputs may differ from the frozen archive; the adapter is not evidence that an external provider has reproduced these outcomes. \texttt{prepare\_panel.py} documents extraction from the private frozen archive; \texttt{analyze.py} and \texttt{make\_figures.py} require the full derived author-side panel to recompute outcome-dependent tables and figures. That panel is retained privately, together with an executed notebook and independent validation checks. No production service calls or new model generations are required when compatible inputs are available. The public release therefore exposes the forecasts, methods, code, and aggregate findings, but is not a fully self-contained outcome-replication dataset. The arXiv source bundle contains only the manuscript, bibliography, and referenced figures.
\clearpage
\section{Framework interpretation and historical modes}\label{app:frameworks}
The following short descriptions explain the platform's configured perspectives; they are not reproductions of every historical prompt. Current framework descriptions cannot certify historical information sets, and persona labels imply no human endorsement. A researcher mode organizes evidence and a thinker mode adds a differently configured reasoning perspective; their effects were not randomized or independently isolated here. Exact archived technical mode strings, including configured creativity/thinking suffixes, and model-version labels remain in the public forecast CSV.
\begin{table}[H]
\centering\small
\begin{tabular}{p{.23\linewidth}p{.24\linewidth}p{.43\linewidth}}
\toprule
Neutral name & Archive label & Interpretation\\
\midrule
Paradigm shift & Elon Musk & Technology adoption, disruption, and execution constraints.\\
Contrarian forensic & Michael Burry & Downside asymmetry, balance-sheet fragility, and accounting risk.\\
Power audit & Niccolo Machiavelli & Incentives, control, counterparties, and distribution of economic rents.\\
Forensic deduction & Sherlock Holmes & Evidence chains, inconsistencies, and competing explanations.\\
Economic machine & Ray Dalio & Macro cycles, liquidity, debt, and regime transmission.\\
Civilizational compute & Superintelligence & Compute and energy constraints across long-term system transitions.\\
Value ownership & Warren Buffett & Durable cash generation, competitive moats, and valuation discipline.\\
\bottomrule
\end{tabular}
\caption{Explanatory framework map. Descriptions summarize configured perspectives, not historical human decisions.}
\end{table}
\section{Additional v2 diagnostics}\label{app:diagnostics}
\begin{table}[H]
\centering\small
\begin{tabular}{lrrr}
\toprule
Comparator & MAE (pp) & Excess (pp) & Direction\\
\midrule
No change & 14.99 & 0.00 & 2.48\%\\
Council mean & 15.15 & 0.16 & 48.38\%\\
Council median & 15.13 & 0.14 & 51.94\%\\
Repeat prior 60-observation return & 25.47 & 10.49 & 42.13\%\\
\bottomrule
\end{tabular}
\caption{Unfitted benchmark diagnostics on 928 outcomes. No-change direction is always flat under the three-category sign rule, not the always-up benchmark. Momentum is an intentionally simple comparator with a different lookback length.}
\end{table}
\begin{table}[H]
\centering\small
\begin{tabular}{lrr}
\toprule
Framework & MAE (pp) & Direction\\
\midrule
Paradigm shift & 15.76 & 49.78\%\\
Contrarian forensic & 17.01 & 47.84\%\\
Power audit & 15.64 & 51.72\%\\
Economic machine & 15.17 & 50.11\%\\
Forensic deduction & 15.93 & 47.95\%\\
Civilizational compute & 15.17 & 49.78\%\\
Value ownership & 15.08 & 49.89\%\\
\bottomrule
\end{tabular}
\caption{Per-framework point forecasts, averaged across retained modes before scoring. These descriptive comparisons are not model-provider rankings.}
\end{table}
\begin{table}[H]
\centering\small
\begin{tabular}{lrrrr}
\toprule
Variable (pp unless noted) & Minimum & Median & 95th pct. & Maximum\\
\midrule
Absolute council forecast & 0.01 & 3.16 & 11.35 & 22.24\\
Realized return & -52.10 & 0.83 & 40.57 & 205.78\\
Council dispersion & 0.22 & 3.87 & 9.26 & 17.83\\
Horizon-scale historical risk & 1.21 & 17.71 & 40.45 & 62.23\\
Horizon (calendar days) & 71.00 & 88.00 & 90.00 & 91.00\\
\bottomrule
\end{tabular}
\caption{Distribution summaries of the 928 primary records.}
\end{table}
\begin{table}[H]
\centering\small
\begin{tabular}{lrrrrr}
\toprule
Subset & Records & Mean risk & Mean spread & Mean momentum & MAE\\
\midrule
Primary balanced & 928 & 20.61 & 4.41 & 4.34 & 15.15\\
Other valid-risk matched & 293 & 27.78 & 5.54 & 7.92 & 17.00\\
\bottomrule
\end{tabular}
\caption{Retained versus excluded valid-risk records; variables and MAE are in pp. The excluded set has 293 records across 141 assets, rather than four observations per asset. Differences describe selection and do not establish its cause.}
\end{table}
The six matched records without valid risk comprise five flagged split-suspect histories and one other unavailable-risk record. They are distinct from the 15 incomplete configuration groups excluded before matching. Primary entry delays are one day for 515 records, two for 202, and three for 211. Daily-history availability and the split heuristic are recorded rather than attributed to delisting without evidence. Asset-class by quartile and selector counts appear in \texttt{results\_v2.json}; the expanded valid-risk set contains 1,221 records for 373 assets.

The original adjusted design has 928 observations and 12 parameters including intercept, three batch indicators, four asset-class indicators, and four continuous terms. Its condition number is 24.88; continuous-term variance inflation factors range from 1.33 to 2.84. The v2 signed-momentum sensitivity adds standardized $\operatorname{asinh}(\text{momentum}/10)$, giving 13 parameters. Its disagreement coefficient is 0.008 [$-0.071$, 0.083] for $\log(1+E)$ ($R^2=0.178$), and $-0.077$ pp [$-0.620$, 0.451] for excess error ($R^2=0.057$). The original model has $R^2=0.174$ for log error and $R^2=0.057$ for excess error; the companion reports all coefficients. Sparse asset classes and resamples without a class limit interpretation of class coefficients; these are not validated error predictors.

Empirical coverage of the council mean plus/minus one, two, and three times framework spread is 21.34\%, 38.25\%, and 53.77\%. These are observed coverages of specified intervals, not nominal Gaussian confidence levels, and no calibrated predictive distribution is inferred. Multiple exploratory analyses increase the opportunity for favorable patterns. All tested thresholds and selectors are reported; none is claimed to have been chosen on an untouched test set.

\begin{table}[H]
\centering\small
\begin{tabular}{lrrr}
\toprule
Omitted batch & Mean-sign gap & Modal gap & Excess-error Q4--Q1\\
\midrule
3 & 13.55 & 2.86 & -0.59\\
4 & 7.07 & -1.89 & -0.35\\
5 & 6.56 & -4.37 & -0.59\\
6 & 9.41 & -0.62 & -0.53\\
\bottomrule
\end{tabular}
\caption{Leave-one-batch-out point sensitivities. Accuracy gaps are high-minus-lower agreement in percentage points; excess-error gaps are pp. These overlapping subsets are not independent validation samples.}
\end{table}
\paragraph{Reconstruction boundary.} \texttt{reconstruct\_panel.py} accepts the public original-anchor forecasts, cohort metadata, and a compatible external market-input table. The recipe specifies anchor-to-entry conversion, outcome dates, price-only native-currency returns, pre-issue historical windows, and split handling. The public edition includes a synthetic fixture and a schema, not the licensed input rows. The adapter was checked against the private frozen derived panel; this is author-side validation, not independent external replication. Without compatible original anchor prices, histories, and corporate-action conventions, substituting a different price feed can change the estimand and results.

\paragraph{Version record.} V2 adds modal voting, ties and threshold diagnostics, resampled-rank selection, scale-aware losses, dependence matrices, period sensitivities, cohort diagnostics, and an external-input adapter. The outcome cutoff, eligible configuration set, and primary cohort remain unchanged. Original fixed-membership calculations retain seed 20261009; resampled-rank and signed-momentum diagnostics use seed 20261014. No new model invocation, retrospective prompt regeneration, or extension to the immature Batch 7 is part of this revision.

\section{Inference and reporting details}

Sector/class-block intervals are [3.95, 16.14] points for the mean-sign high-minus-lower accuracy gap and [$-5.92$, 5.93] for the modal-direction gap. The excess-error Q4--Q1 interval is [$-2.32$, 0.95] pp. All 5,000 block draws have nonempty comparison groups. These are a dependence sensitivity, not a correction that makes four common market periods independent.

For the original log-error and excess-error regressions, residual standard errors are 0.889 log units and 5.088 pp. Partial $R^2$ for disagreement is 0.000022 and 0.000134, conditional on the specified other covariates; neither is a causal importance measure. In the raw-error sensitivity the standardized disagreement coefficient is +0.219 pp [$-1.786$, 2.500], with $R^2=0.160$ and residual standard error 16.974 pp. Coefficient intervals remain conditional on observed periods.

Council-minus-no-change RMSE is +1.135, +0.943, +0.598, and +0.511 pp across the four disagreement quartiles. The Q4--Q1 contrast interval is [$-2.05$, 1.04] pp, computed from paired resamples of squared errors followed by square roots. The lowest-quartile difference has an interval [0.061, 1.977] pp, reinforcing that small raw loss in this subset is not the same as beating its baseline under every loss.

The companion's analysis inventory enumerates executed comparisons. V2 reports 16 sign-band/cutoff combinations, each with higher/lower agreement groups, and 80 selector/coverage combinations. They complement the original cohort, spread, aggregation, and risk-window sensitivities; they are not independent replications or a multiplicity-controlled confirmatory family. The study developed through exploratory drafting from one frozen outcome vintage. A prospective specification log was not maintained, so the inventory should not be interpreted as proof that no other exploratory choice was considered.

\begin{table}[H]\centering\small

\begin{tabular}{lrrrr}\toprule

Batch & $D$--risk & Risk--error & $D$--error & $D$--excess error\\\midrule

3 & 0.562 & 0.482 & 0.214 & -0.026\\

4 & 0.641 & 0.420 & 0.286 & -0.091\\

5 & 0.689 & 0.307 & 0.236 & -0.025\\

6 & 0.595 & 0.371 & 0.321 & 0.009\\

\bottomrule\end{tabular}

\caption{Within-batch Spearman coefficients in the primary panel. $D$ is numerical disagreement. These rank correlations are descriptive, not causally identified components of uncertainty.}\end{table}

Exact issue, entry, target, and observed dates for every matched record are in public \texttt{cohort.csv}, allowing the timing figure to be checked without price rows. Counts of all nine predicted/outcome sign combinations are in \texttt{results\_v2.json}, including the flat class and modal ties. Regression coefficients and intervals, class/selector composition, risk-window overlap, and expanded/equity-only directional sensitivities are also reported there. Point-estimate tables in the main text are paired with uncertainty plots or released interval tables; no omitted interval is being treated as zero uncertainty.


\clearpage
{\small\begin{thebibliography}{10}
\bibitem{lahiri2010} K. Lahiri and X. Sheng. Measuring forecast uncertainty by disagreement: The missing link. \emph{Journal of Applied Econometrics}, 25:514--538, 2010. \href{https://doi.org/10.1002/jae.1167}{doi:10.1002/jae.1167}.
\bibitem{glas2020} A. Glas. Five dimensions of the uncertainty--disagreement linkage. \emph{International Journal of Forecasting}, 36(2):607--627, 2020. \href{https://doi.org/10.1016/j.ijforecast.2019.07.010}{doi:10.1016/j.ijforecast.2019.07.010}.
\bibitem{lakshminarayanan2017} B. Lakshminarayanan, A. Pritzel, and C. Blundell. Simple and scalable predictive uncertainty estimation using deep ensembles. \emph{Advances in Neural Information Processing Systems}, 30, 2017. \href{https://proceedings.neurips.cc/paper/2017/hash/9ef2ed4b7fd2c810847ffa5fa85bce38-Abstract.html}{NeurIPS proceedings}.
\bibitem{xiao2025} Q. Xiao, D. Bhattacharjya, B. Ganesan, R. Marinescu, K. Mirylenka, N. H. Pham, M. Glass, and J. Lee. The consistency hypothesis in uncertainty quantification for large language models. \emph{Proceedings of the Forty-first Conference on Uncertainty in Artificial Intelligence}, PMLR 286:4636--4651, 2025. \href{https://proceedings.mlr.press/v286/xiao25a.html}{PMLR proceedings}.
\bibitem{karger2024} E. Karger, H. Bastani, Chen Yueh-Han, Z. Jacobs, D. Halawi, F. Zhang, and P. E. Tetlock. ForecastBench: A dynamic benchmark of AI forecasting capabilities. \emph{arXiv:2409.19839}, 2024, revised 2025. \url{https://arxiv.org/abs/2409.19839}.
\bibitem{elyaniv2010} R. El-Yaniv and Y. Wiener. On the foundations of noise-free selective classification. \emph{Journal of Machine Learning Research}, 11:1605--1641, 2010. \href{https://www.jmlr.org/papers/v11/el-yaniv10a.html}{JMLR article}.
\bibitem{gneiting2011} T. Gneiting. Making and evaluating point forecasts. \emph{Journal of the American Statistical Association}, 106(494):746--762, 2011. \href{https://doi.org/10.1198/jasa.2011.r10138}{doi:10.1198/jasa.2011.r10138}.
\bibitem{ramdowar2026} R. Ramdowar. When models judge models: How source identity and disagreement shape multi-agent financial forecasts. iPulse AI research artifacts, 2026. \href{https://doi.org/10.5281/zenodo.23083356}{doi:10.5281/zenodo.23083356}.
\bibitem{christoffersen2006} P. F. Christoffersen and F. X. Diebold. Financial asset returns, direction-of-change forecasting, and volatility dynamics. \emph{Management Science}, 52(8):1273--1287, 2006. \href{https://doi.org/10.1287/mnsc.1060.0520}{doi:10.1287/mnsc.1060.0520}.
\bibitem{hyndman2006} R. J. Hyndman and A. B. Koehler. Another look at measures of forecast accuracy. \emph{International Journal of Forecasting}, 22(4):679--688, 2006. \href{https://doi.org/10.1016/j.ijforecast.2006.03.001}{doi:10.1016/j.ijforecast.2006.03.001}.
\end{thebibliography}
}
\end{document}